# Spherical "Top-Hat" Collapse in a Modified Chaplygin Gas Dominated Universe

**S. Karbasi** [(1)] and **H. Razmi** [(2)]

***Department of Physics, the University of Qom, 3716146611, Qom, I. R. Iran***

(1) s.karbasi@stu.qom.ac.ir

(2) razmi@qom.ac.ir & razmiha@hotmail.com

## Abstract

Considering perturbation growth in spherical Top-Hat model of structure formation in a generalized Chaplygin gas dominated universe, we want to study this scenario with modified Chaplygin gas obeying an equation of state $p = A\rho - B/\rho^{\alpha}$ model. Different parameters of this scenario for positive and negative values of $A$ are computed. The evolution of background and collapsed region parameters are found for different cases. The stability of the model and the collapse time rate are considered in different cases. The turn-around redshifts for different values of α are computed; the results are in relatively good agreement with current observational data.

PACS: 98.80.-k, 98.80.Es, 95.35.+d, 95.36.+x

KEYWORDS: cosmology, Chaplygin gas, spherical collapse, structure formation, Top-Hat model

## 1. Introduction

Recent cosmological data and information from modern observations and study of the cosmic microwave background radiation (CMBR) [1-2], large scale structure [3-4], and type Ia supernovae [5-6], display the universe endures an accelerated expansion. In general relativity, an unknown energy component is introduced to explain this accelerated expansion of the universe. In cosmology, the simplest candidate for this unknown energy, later on named as dark energy (DE), is the cosmological constant [7-8]. On the other hand, astronomical observations of galaxy clusters indicate the universe is dominated by non-baryonic dark matter (DM) [9]. Ordinary DM is assumed as a dust similar to a fluid with no pressure and DE is modeled as a fluid with negative pressure acting against gravitational attraction and collapse. The simplest model which incorporates the DM and DE components is the so-called $\Lambda$CDM model, in which the cosmological constant ($\Lambda$) plays the role of DE and particles with a negligible free streaming length have the role of Cold Dark Matter (CDM). Based on the present knowledge, DE and DM account about 95% of the energy density of the universe today.

Among a number of different considerations and models, one of the conjectures is that DE and DM can be unified using Chaplygin gas (CG) model [10]. CG obeys an equation of state $p = -B/\rho$, $B > 0$, where $\rho$ and $p$ are the pressure and energy density respectively and B is a constant [11]. CG model has been extended to the co-called generalized Chaplygin gas (GCG) model with the equation of state $p = -B/\rho^{\alpha}$, $B > 0$, $0 < \alpha < 1$ [12-15]. A simple more generalization, named as the Modified Chaplygin gas (MCG), is achieved by adding a barotropic term [16-17]. In principle, MCG model is used to explain the evolution of the universe from early times to the cosmological constant dominated era [18-20]. MCG model can be applied to inflation theory too [21-22]. MCG model includes an initial phase of radiation and is based on the equation of state $p = A\rho - B/\rho^{\alpha}$[23]. For $B = 0$, it reduces to the perfect fluid ($p = A\rho$) model and for $A$=0, it reduces to the GCG ($p = -B/\rho^{\alpha}$) model. The MCG model can be used to explain the radiation era in the early universe; at the later times, MCG behaves as the cosmological constant and can be fitted to $\Lambda$CDM.

Spherical 'Top-Hat' (STH) model is one of the well-known models proposed to explain the large scale structure formation [24] and has already been studied in GCG model [25]. In this paper, we want to study STH model in MCG scenario.

## 2. Spherical 'Top-Hat' collapse of Chaplygin gas

The spherical collapse describes the evolution of an initial spherical perturbation in a homogeneous background. Spherical 'Top-Hat' collapse model describes the evolution of a closed structure like a homogeneous universe with positive curvature in FRW model. In this model, the equations for background evolution are

$$\dot{\rho} = -3h(\rho + p) \tag{1}$$

$$\frac{\ddot{a}}{a} = -\frac{4\pi G}{3}\sum_j (\rho_j + 3p_j) \tag{2},$$

and the basic equations in the perturbed region are

$$\dot{\rho}_c = -3h(\rho_c + p_c) \tag{3}$$

$$\frac{\ddot{r}}{r} = -\frac{4\pi G}{3}\sum_j (\rho_{cj} + 3p_{cj}) \tag{4},$$

where $\rho_c = \rho + \delta\rho$, $p_c = p + \delta p$ are the perturbed density and pressure quantities, and $h$ is related to the Hubble parameter as in the following [26]

$$h = H + \frac{\theta}{3a} \tag{5}$$

in which $\theta \equiv \vec{\nabla}.\vec{v}$ is the divergence of the peculiar velocity $\vec{v}$.

Using the time derivatives of density contrast $\delta_j = (\delta\rho/\rho)_j$ and $\theta$

$$\dot{\delta}_j = -3H(c_{effj}^2 - w_j)\delta_j - [1 + w_j + (1 + c_{effj}^2)\delta_j]\frac{\theta}{a} \tag{6}$$

$$\dot{\theta} = -H\theta - \frac{\theta^2}{3a} - 4\pi G a \sum_j \rho_j \delta_j (1 + 3c_{effj}^2) \tag{7},$$

the dynamical evolution equations of density contrast $\delta_j$ and $\theta$ are found as

$$\delta'_j = -\frac{3}{a}(c_{effj}^2 - w_j)\delta_j - [1 + w_j + (1 + c_{effj}^2)\delta_j]\frac{\theta}{Ha^2} \tag{8}$$

$$\theta' = -\frac{\theta}{a} - \frac{\theta^2}{3Ha^2} - \frac{3H}{2}\sum_j \Omega_j \delta_j \left(1 + 3c_{effj}^2\right) \quad (9),$$

where prime denotes the derivative with respect to scale factor, and $c_{effj}^2 = (\delta p/\delta\rho)_j$is the square of the effective sound speed,$\Omega_j = \frac{8\pi G}{3H^2}\rho_j$ is the relative density parameter, and the index *j* refers to MCG and baryonic matter.

The Hubble parameter is [27]

$$H(t) = H_0\sqrt{(1-\Omega_b-\Omega_r-\Omega_k)[\bar{c}+(1-\bar{c})a^{-3(1+\alpha)(1+A)}]^{\frac{1}{1+\alpha}} + \frac{\Omega_r}{a^4} + \frac{\Omega_b}{a^3} + \frac{\Omega_k}{a^2}} \quad (10),$$

where $\Omega_b$ , $\Omega_r$, and $\Omega_k$denote the dimensionless baryonic matter, radiation, and curvature density respectively. Since the radiation density doesn't any considerable effect in the structure formation and the universe is nearly flat, $\Omega_r$ and $\Omega_k$can be neglected.

## 3. Modified Chaplygin gas

Let begin with the equation of state in MCG model

$$p = A\rho - \frac{B}{\rho^\alpha} \quad (11),$$

And the equation of energy density

$$\rho = \rho_0(\bar{c} + (1-\bar{c})a^{-3(1+\alpha)(1+A)})^{\frac{1}{1+\alpha}} \quad (12)$$

where, $\bar{c} = \frac{B}{(1+A)\rho_0^{1+\alpha}}$, $\rho_0$ is the density at the present time, $a$ is the cosmic scale factor, and $1 + z = \frac{1}{a}$(i.e. $a_0 = 1$).

It can be easily shown that the equation of state parameter $w = p/\rho$ is given by

$$w = A - \frac{B}{\rho^{1+\alpha}} = A - \frac{\bar{c}(1+A)}{\bar{c}+(1-\bar{c})a^{-3(1+\alpha)(1+A)}} \quad (13)$$

The state parameter relative to the collapsing region ($w_c$) is given by

$$w_c = \frac{p+\delta p}{\rho+\delta\rho} = \frac{w}{1+\delta} + c_{effj}^2\frac{\delta}{1+\delta} \quad (14)$$

The coefficient factor $c_{eff}^2$is present only if the perturbation exists; so, $c_{eff}^2 = \frac{\delta p}{\delta \rho}$by which one can determine the dynamical behavior of the perturbed region.

Using $\rho_c = \rho(1+\delta)$, the effective sound speed is found in terms of background parameter $w$ and collapsed region parameter $\delta$ as in the following

$$c_{eff_j}^2 = \left(\frac{1}{\delta}\right)\left(A(1+\delta) - \frac{\bar{c}(1+A)(1+\delta)^{-\alpha}}{\bar{c}+(1-\bar{c})a^{-3(1+\alpha)(1+A)}} - w\right) \quad (16)$$

The adiabatic sound speed is

$$c_s^2 = \frac{\partial P}{\partial \rho} = A + \alpha \frac{\bar{c}(1+A)}{\bar{c}+(1-\bar{c})a^{-3(1+\alpha)(1+A)}} \quad (17)$$

In the next section, the above parameters are computed numerically (graphically) for different values of $\alpha$ in two cases *A*> 0 and *A*< 0.

## 4. The method and the result

Let consider non-linear evolution of the baryon and MCG perturbations in the STH model for two cases *A*<0 and *A*>0. Integrations of the equations (8) and (9) are done using the fourth order Runge-Kutta algorithm in FORTRAN software for *z* = 1000 (at recombination), to *z* = 0 (the present time). The background is qualified by a flat FRW universe. The values of density parameters for MCG and baryons are assumed to be $\Omega_{mcg}$= 0.95 and $\Omega_b$= 0.05 respectively; with a Hubble constant of the value of $H_0$= 72km/(sMpc). The parameter $\bar{c}$ is fixed at $\bar{c} = 0.73$ for different values of α. Let now consider two arbitrary values for *A* (*A* = 0.002 & *A*= –0.05). It should be noticed that these two values aren't preferred specific values except they are typical positive and negative possible values for *A* in the range *-1*<*A*<1 . The initial conditions for the system are considered as $\delta_b(z=1000) = 10^{-5}$ and $\theta = 0$ [25]. For the case *A* = 0.002, $\delta_{\boldsymbol{MCG}}(z=1000) = 3.5 \times 10^{-3}$; and, for *A* = –0.05, $\delta_{Mcg}(z=1000) = 3.5 \times 10^{-2}$. Obviously, for *A*<0, the initial perturbation, to form the structure, should have a greater value; because the initial pressure is negative and acts against gravitational collapse.

For $A$>0, the initial pressure is positive and, with the help of gravity, can form the structure by lesser values of perturbation. Baryons can be assumed as dust; $p_b = w_b = c_{effb}^2 = c_s^2 = 0$ [25].

The evolutions of perturbations are shown in figures 1and 2 for MCG and baryonic matter respectively for different values of α. As is seen and expected, the perturbation growth for $A$ = 0.002 is quicker than $A$ = – 0.05. The parameter α has an important role in the results; higher values of this parameter correspond to faster collapses (quick growth of the perturbations).

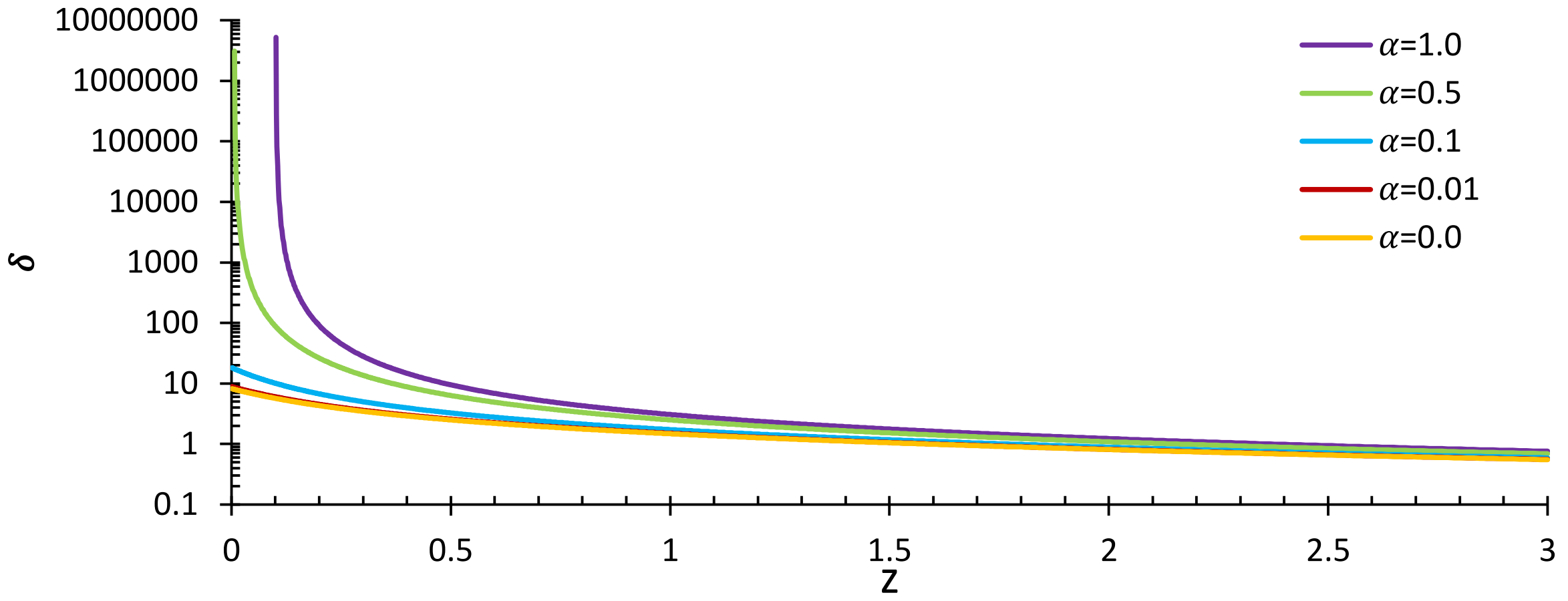

Fig.1.a. The MCG perturbation growth in spherical Top-Hat model ($A$ = 0.002).

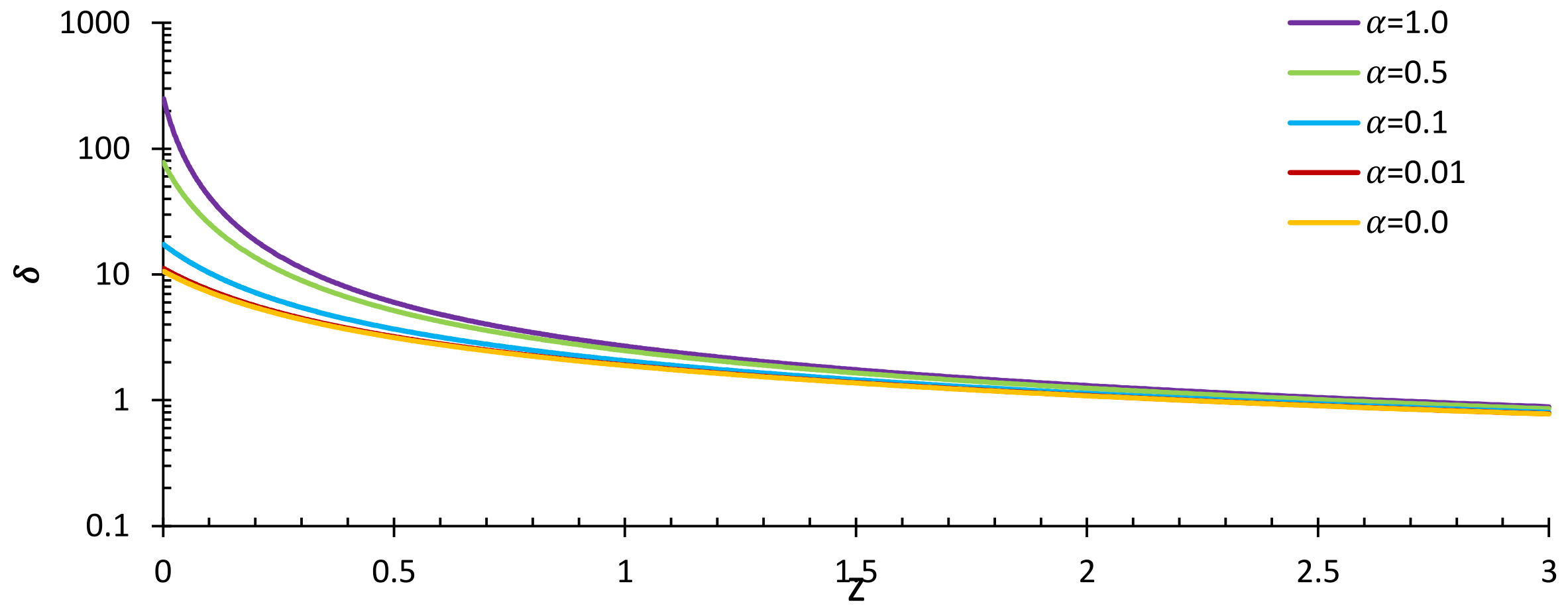

Fig.1.b. The MCG perturbation growth in spherical Top-Hat model ($A$ = – 0.05).

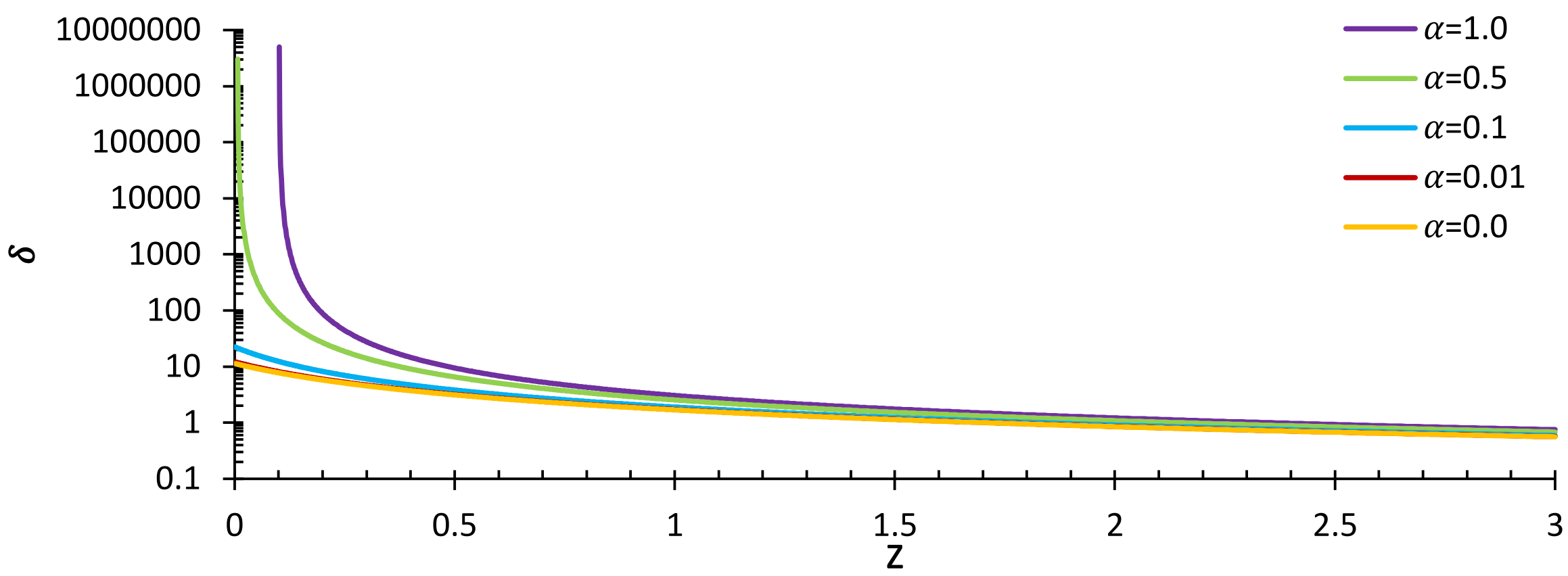

Fig.2.a. Baryonic matter perturbation growth in spherical Top-Hat model ($A$ = 0.002).

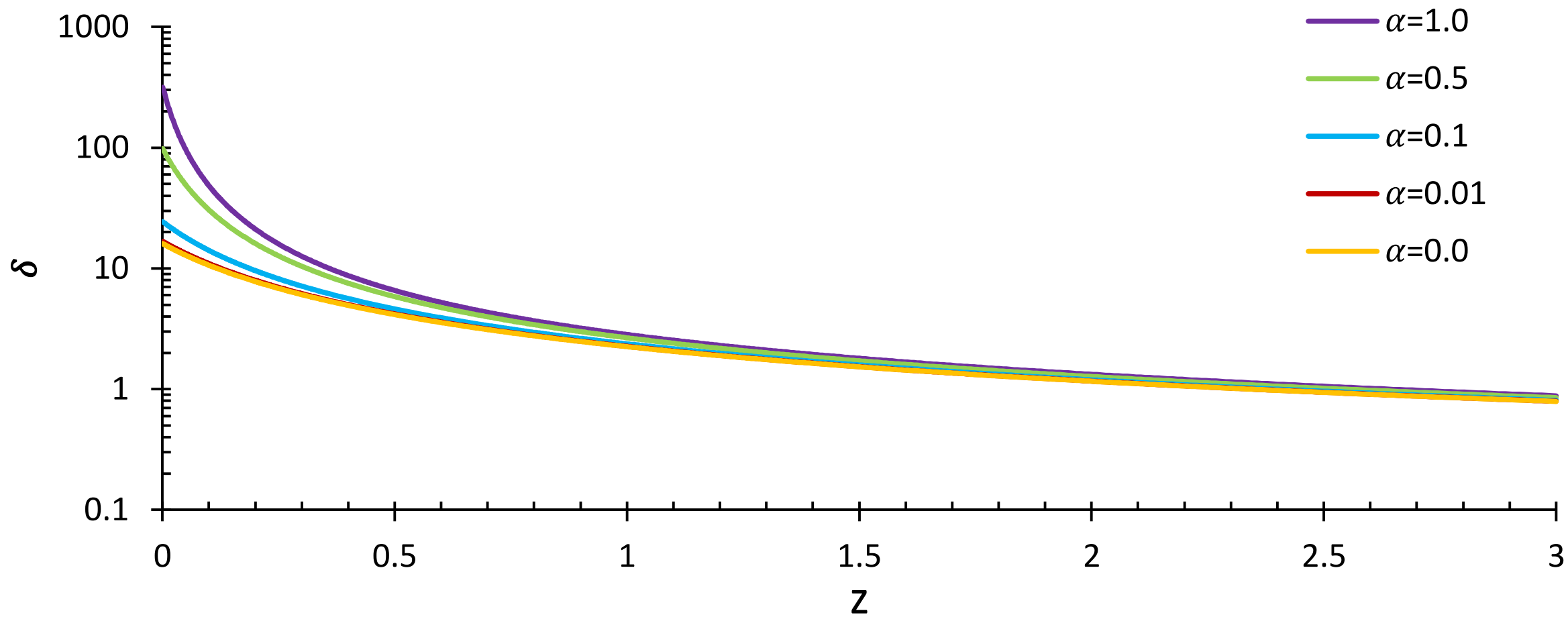

Fig.2.b.Baryonic matter perturbation growth in spherical Top-Hat model($A$ = – 0.05).

The background state parameter $w$ and the parameter relative to the collapsing region$w_c$, are represented in figures 3 and 4.

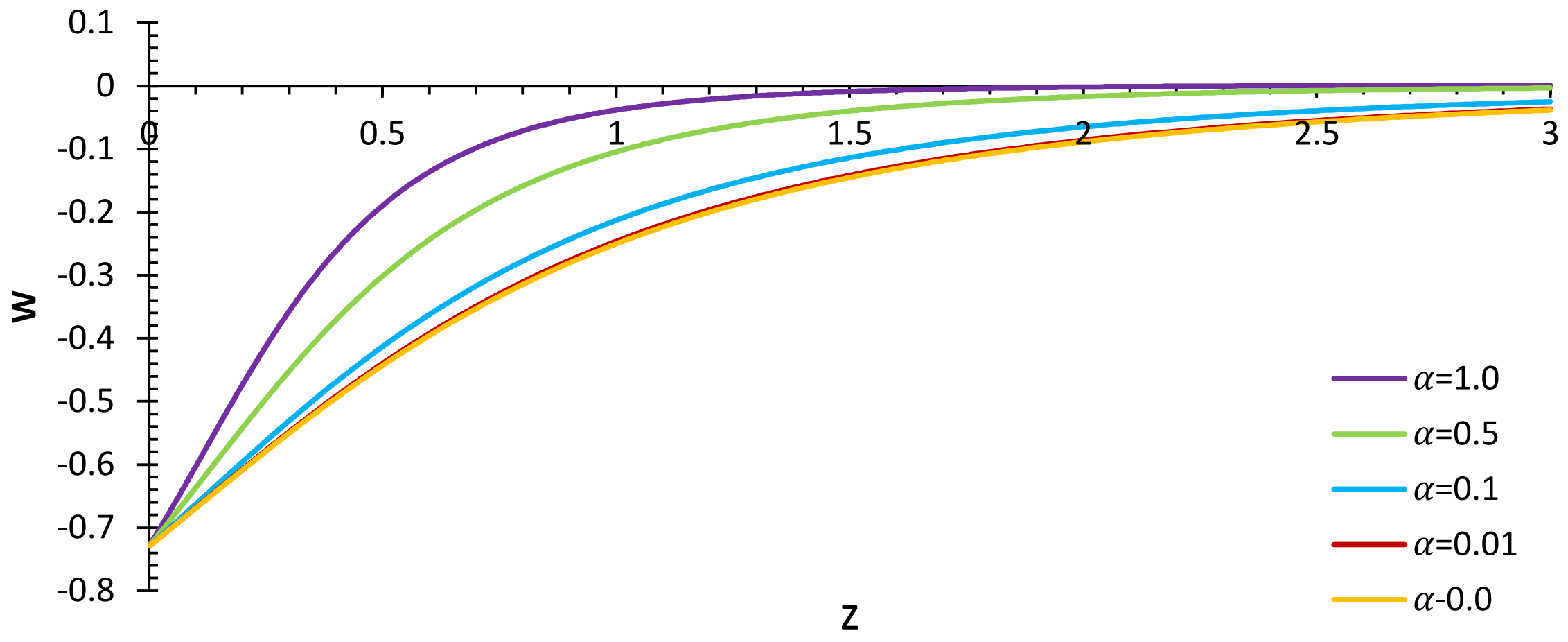

Fig.3.a. Evolution of $w_c$ with respect to redshift in MCG model for different values of $\alpha$ ($A$ = 0.002).

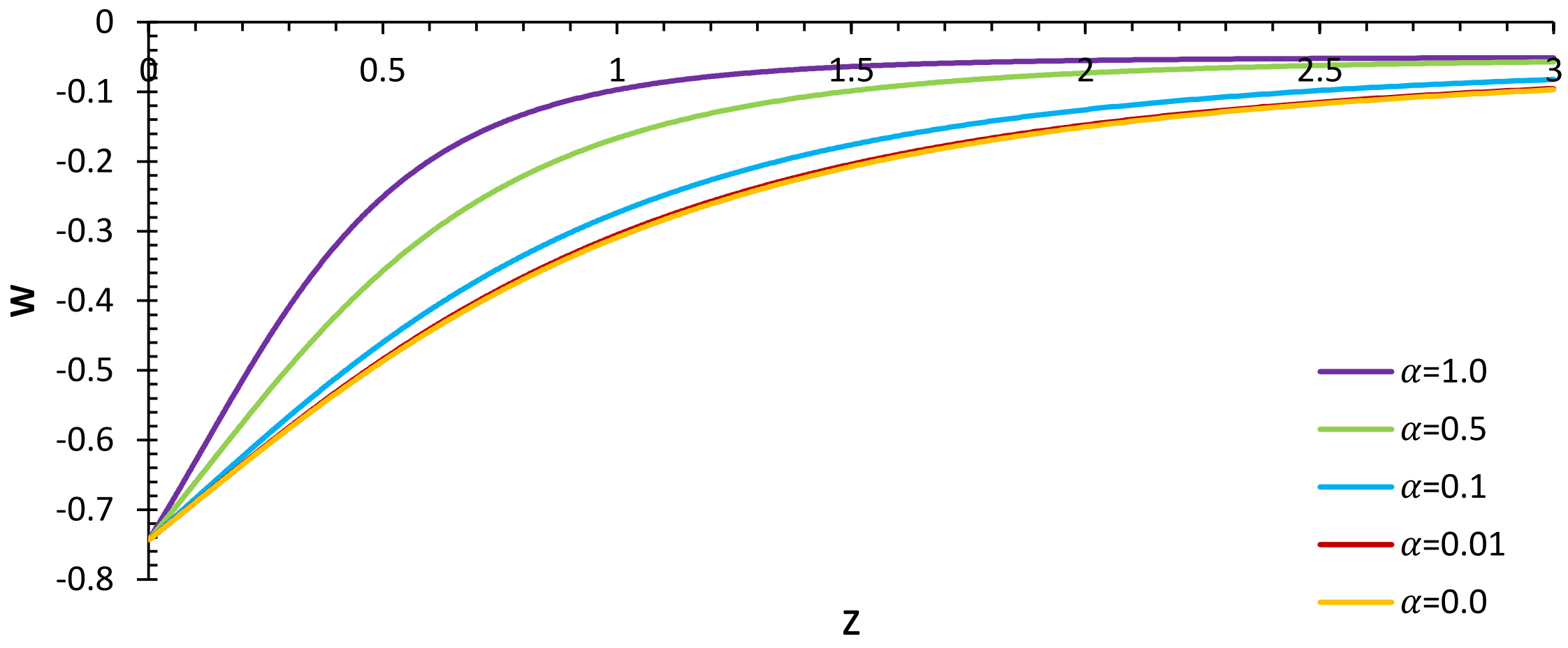

Fig.3.b. Evolution of $w_c$ with respect to redshift in MCG model for different values of $\alpha$ ($A$ = – 0.05).

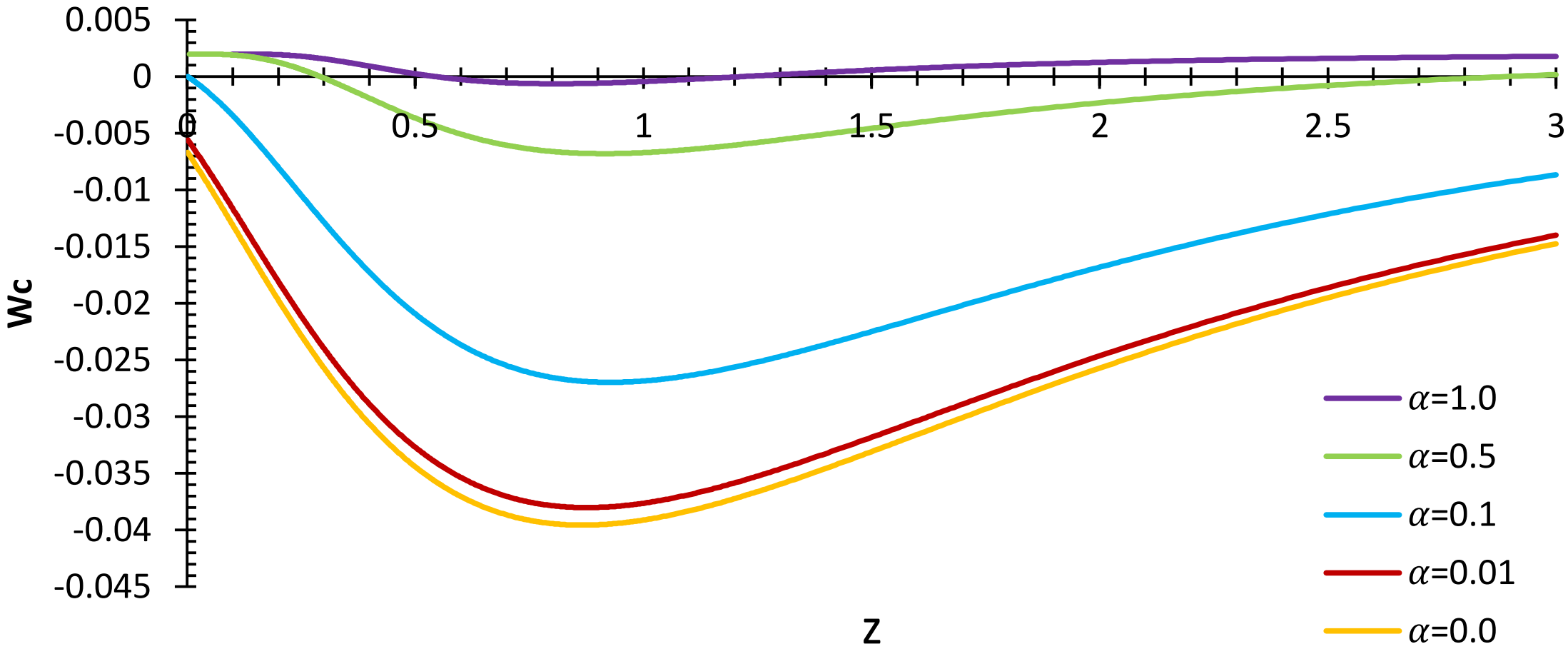

Fig.4.a. Evolution of $w_c$ with respect to redshift in MCG model for different values of $\alpha$ ($A$ = 0.002).

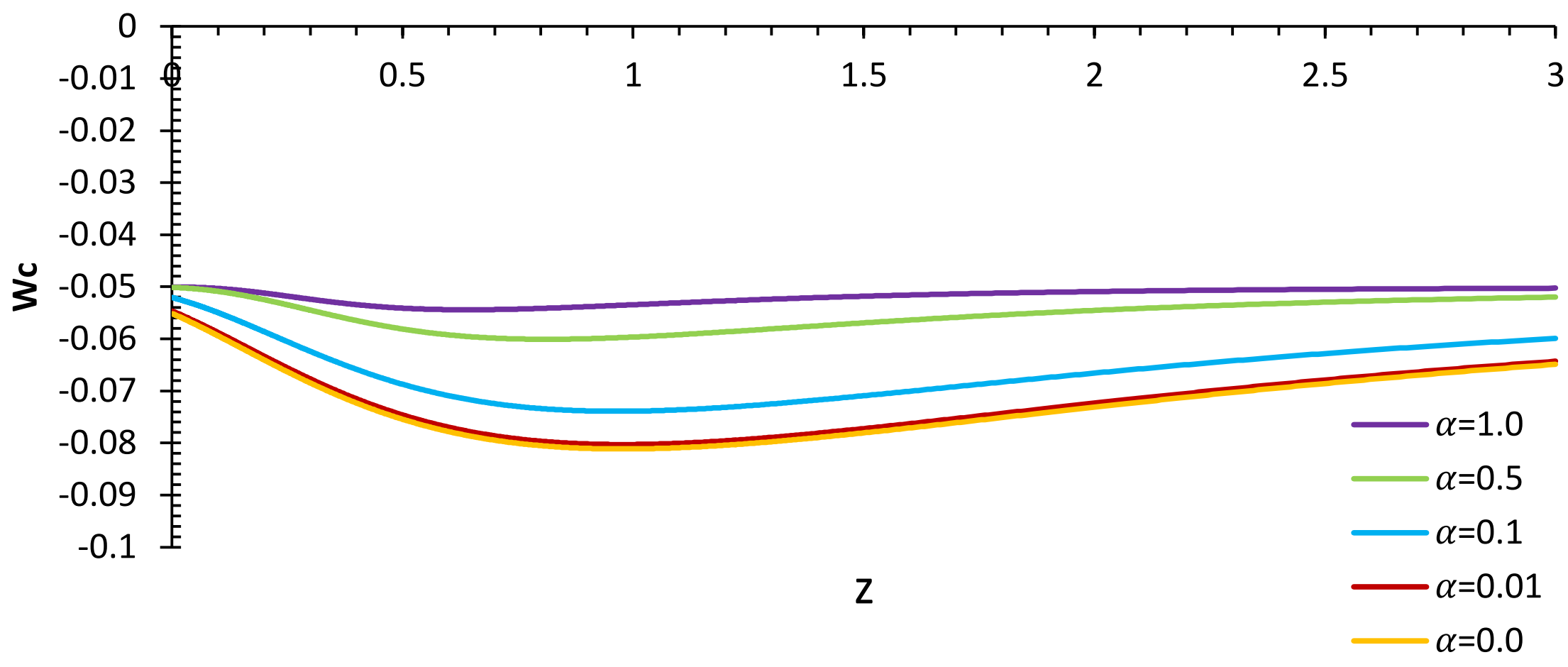

Fig.4.b.Evolution of $w_c$ with respect to redshift in MCG model for different values of $\alpha$ ($A$ = – 0.05).

For small values of the scale factor, using (13), $w \longrightarrow A$; thus, in the early universe, $w$ has negative values for $A< 0$.

To find $w_c$, both of $A$ and $\alpha$ have important roles. Generally, $w_c$ is negative; but, it is positive for special case ($A$ = 0.002 and $\alpha$ = 1.0, 0.5)!

As is seen in the fig.5, $c_{eff}^2$ depends on $A$ intensely. Sign of $c_{eff}^2$ is changed with different values of $A$. Based on the equation(16),the larger values of $\delta$ result in smaller values of $c_{eff}^2$; this behavior is expected in a MCG clump.

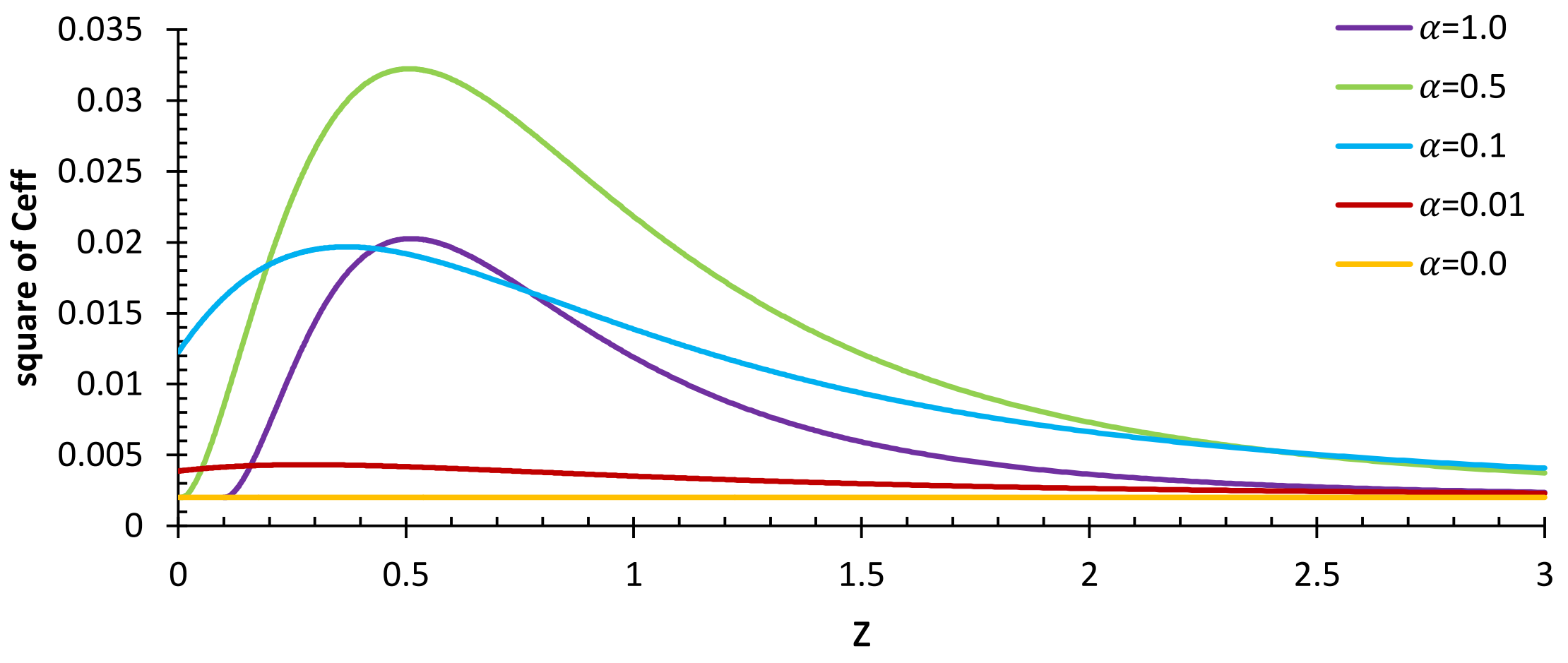

Fig.5.a. Evolution of $c_{eff}^2$ with respect to redshift in MCG model for different values of $\alpha$ ($A$ = 0.002).

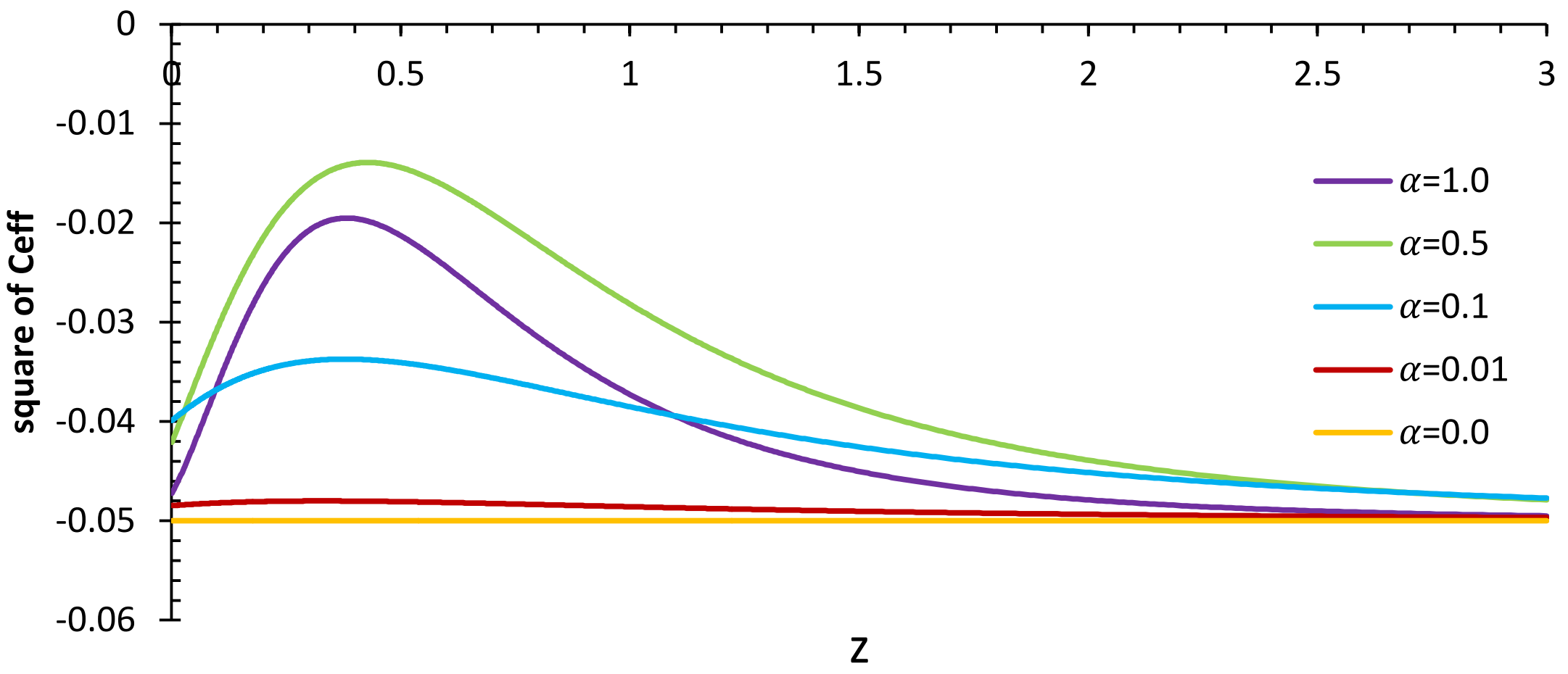

Fig.5.b.Evolution of $c_{eff}^2$ with respect to redshift in MCG model for different values of $\alpha$ ($A$ = – 0.05).

As is seen,$c_{eff}^2$ is negative for negative values of $A$; this doesn't cause any difficulty, because the physical measurable quantity is the adiabatic sound speed whose square value is $c_s^2 = \frac{\partial p}{\partial \rho}$ and its behavior has been shown in fig.6. For $A$ >0, the system is always stable; but for $A$< 0the stability depends on the magnitudes of $c_s^2$and $\alpha$ relative to each other. In the case of $A$< 0, for $\alpha$ = 0.0, 0.01, the system is always instable; but for $\alpha$ = 0.1, 0.5, 1.0, it is stable at present time. Of course, it should be mentioned that the instability of the MCG fluid at high z, (i.e. in the early universe) can be useful in the structure formation [28].

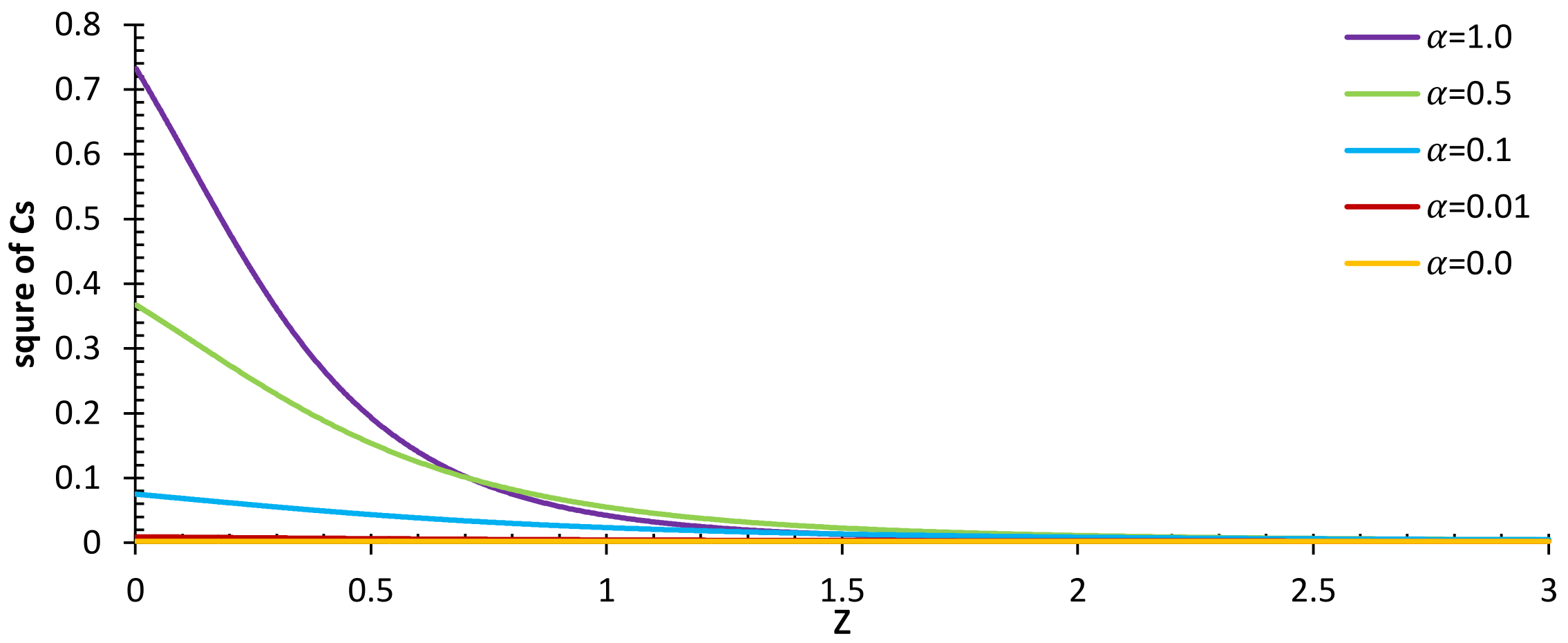

Fig.6.a. Evolution of $c_s^2$ with respect to redshift in MCG model for different values of $\alpha$ ($A$ = 0.002).

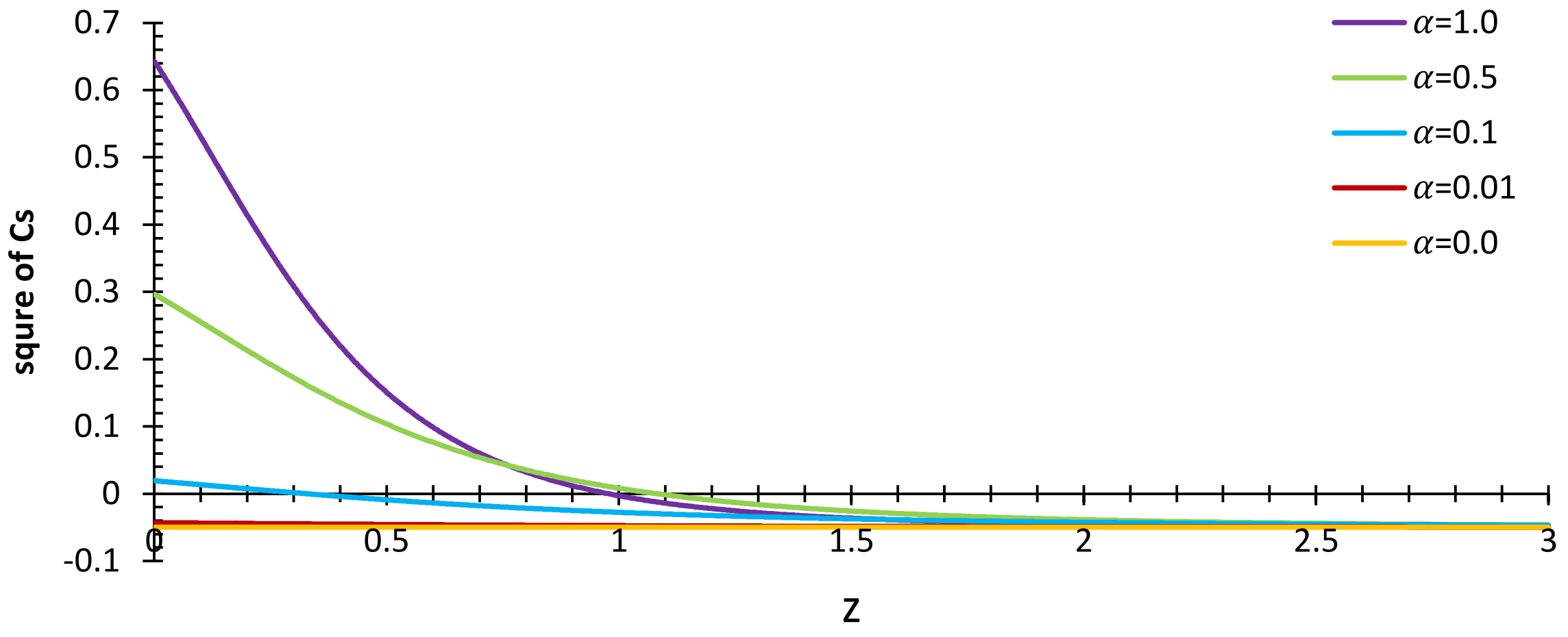

Fig.6.b. Evolution of $c_s^2$ with respect to redshift in MCG model for different values of($A$ = – 0.05).

Finally, we consider the evolution of the rate of collapsed region, h, with respect to z, in figure 7.

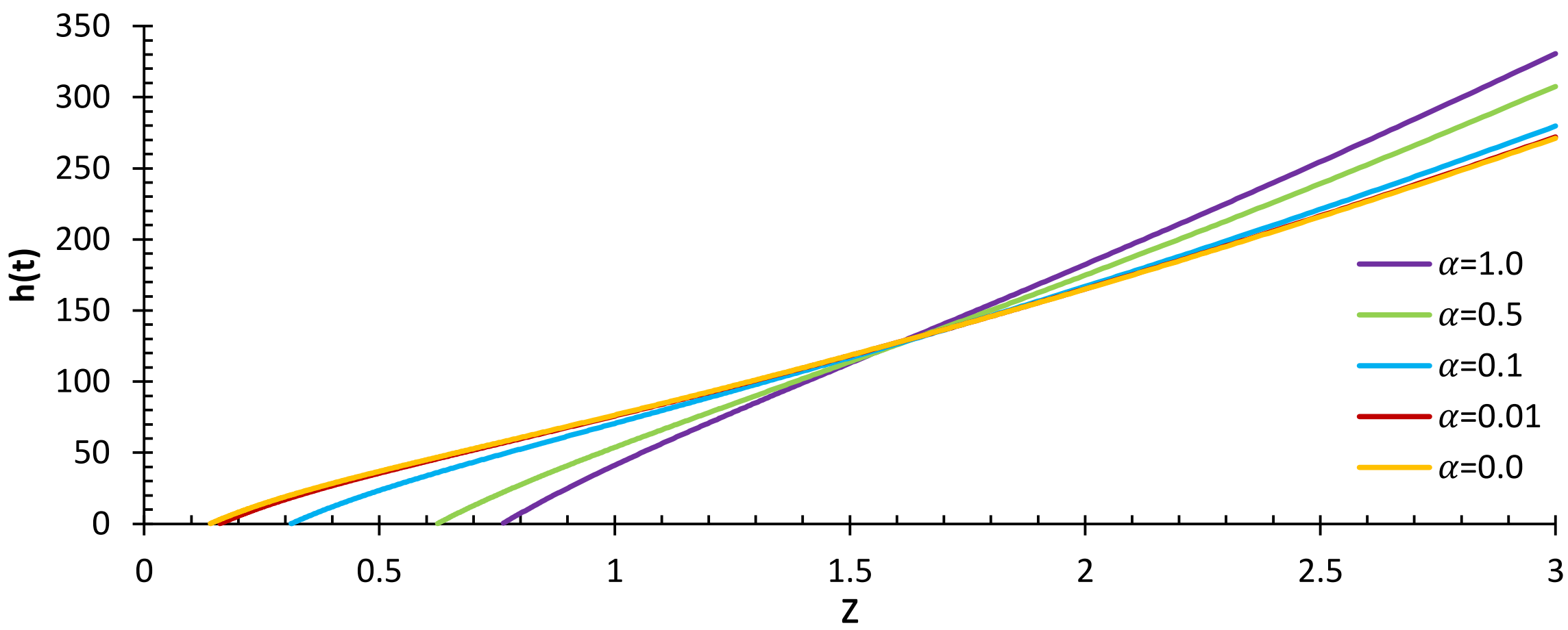

Fig.7.a. Evolution of collapsed region expansion rate, h, with respect to z for $A$ = 0.002.

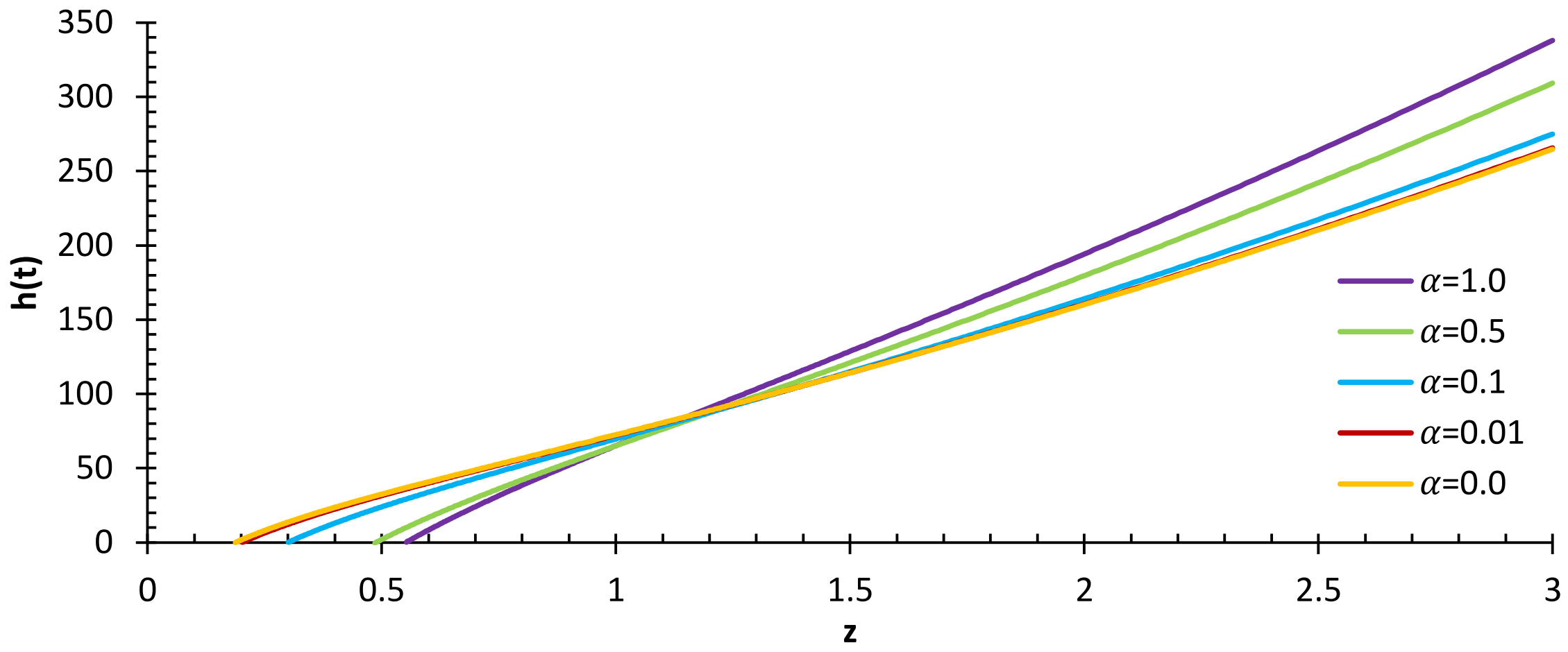

Fig.7.b.Evolution of collapsed region expansion rate, h, with respect to z for $A$ = – 0.05.

As is seen, the turn around redshift $z_{ta}$ is directly proportional to $\alpha$.

In MCG scenario, the constant $\alpha$ should satisfy 0<$\alpha$<1 to justify the accelerated expansion of the universe; but, in GCG scenario, because of both theoretical considerations and some observational data restrictions, $\alpha$ should have a negative value [29-32]. For $\alpha$<0, the state equation corresponds to a decelerated expansion at the present time epoch or a closed universe; but, it seems it is still possible to construct a GCG model based on a more fundamental theory without suffering from these difficulties [32].

Although in this paper, we have studied MCG scenario, let work with a negative value of $\alpha$ too. Considering a value of $\alpha$ = -0.089 which has been already considered in e.g. [32], we can find a suitable MCG scenario with an acceptable collapse time for arbitrary positive value of $A$=0.01 (see the following figures and curves):

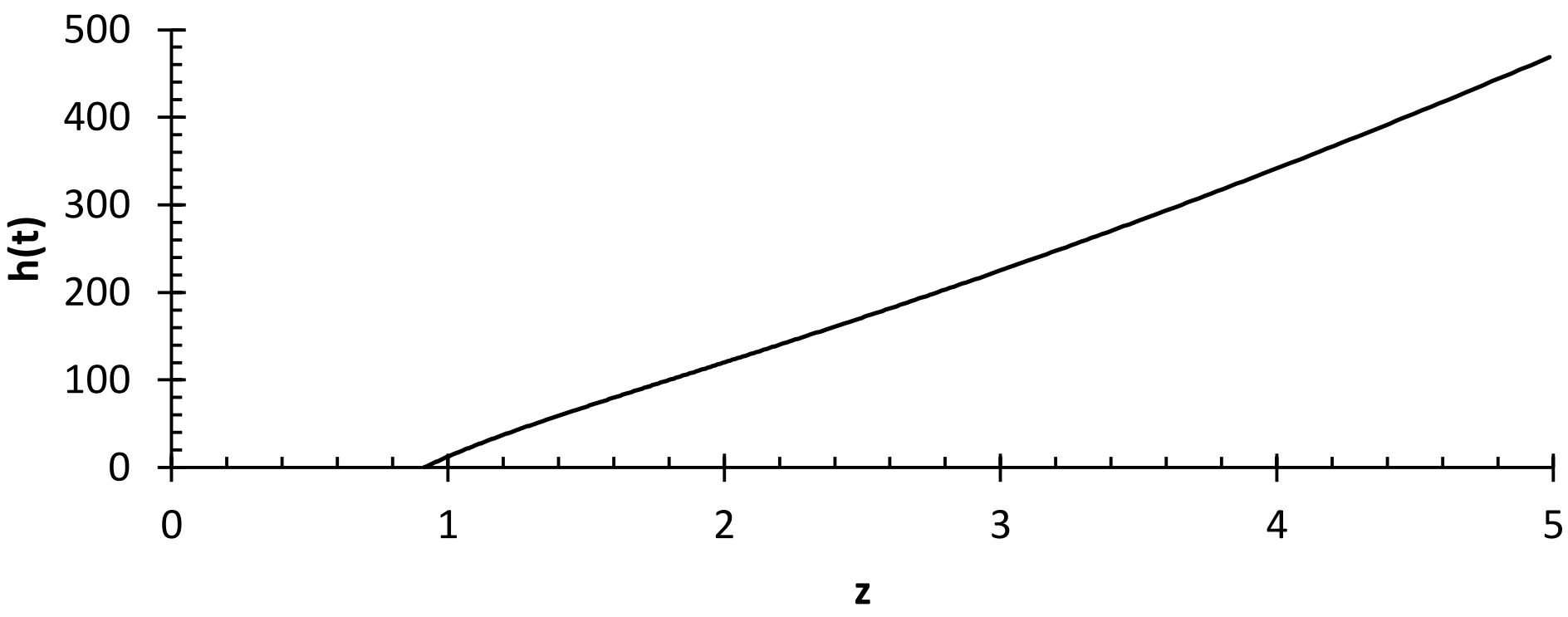

Fig.8.a. Evolution of collapsed region expansion rate, h, with respect to z for $A$ = 0.01, $\alpha$= -0.089.

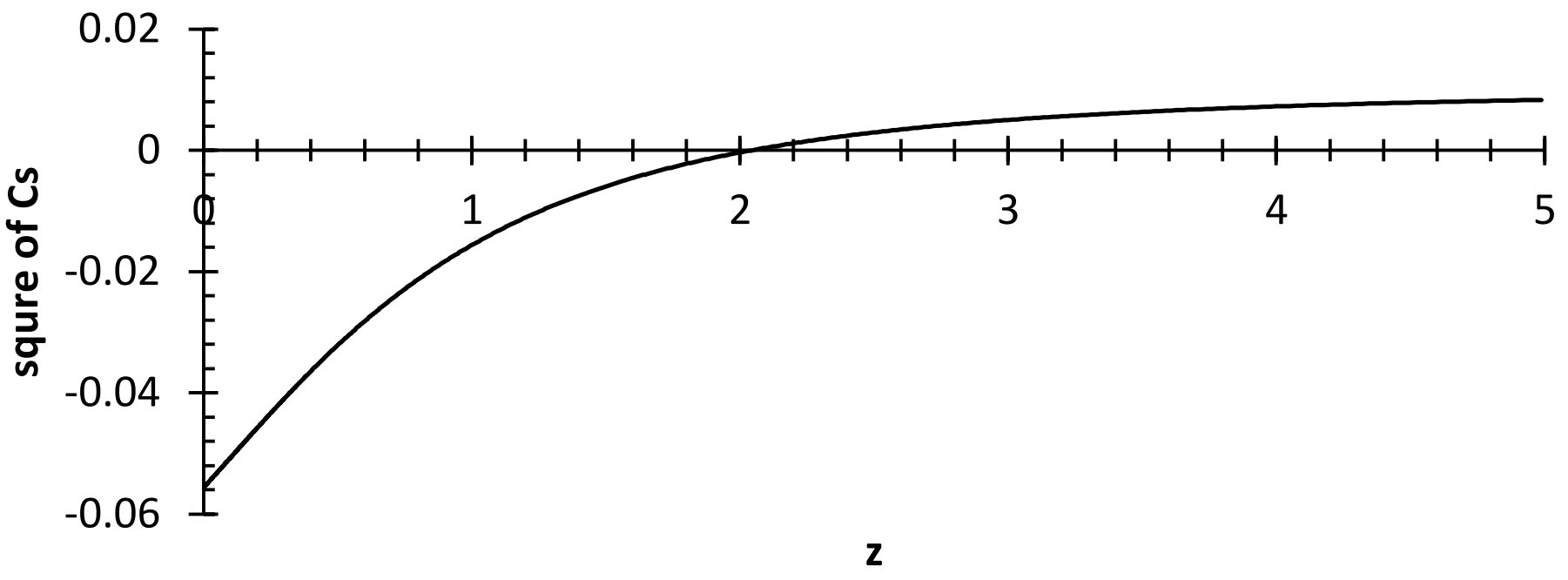

Fig.8.b. Evolution of $c_s^2$with respect to redshift in MCG model for $A$ = 0.01, $\alpha$= -0.089.

For negative values of $\alpha$, the adiabatic sound velocity decreases with increasing time while for positive values of $\alpha$ one dealt with a direct proportionality of the adiabatic sound velocity and time.

## 5. Conclusion

Considering perturbation growth in a GCG-dominated universe in STH model [25] and the study of MCG parameters [28], here, we have studied perturbation growth in MCG-dominated universe in STH profile. The evolution of background parameters $(w, c_s^2)$ and collapsed region parameters $(w_c, c_{eff}^2)$ have been found for different cases. As was seen in figures 1 and 2, for positive values of the modification coefficient (the barotropic coefficient) $A$, the present perturbation has very large value (i.e. the density of structure is very large); this can be used in explaining galaxy and star structure formation. For negative values of $A$, the present perturbation has lesser value and thus it can be used in explaining the structure of recently formed cluster (or super cluster) of galaxies. For both of the above cases, the growth of perturbation for baryons has a greater value; this is because baryonic matter has no pressure. According to fig.3, our result is similar to STH in GCG scenario ($A$ = 0) (i.e. the evolution of background in GCG and MCG scenarios is nearly similar to each other).When collapse happens, the structure is under the influence of gravity and there isn't negative pressure; thus, $w_c$should be positive. In fig.4, $w_c$ has been shown and we see it is negative for all except $A$ = 0.002 and $\alpha$ = 1.0, 0.5 values.

In fig.6, we have studied the stability of the system. For $A>0$, the system is stable always for different values of $\alpha$; but, for $A<0$, the stability of the system depends on the values of $A$ and $\alpha$. The stability of the model and the collapse time rate have been studied in different cases.

The turn-around redshifts for different values of α have been computed. Knowing that the large scale structures are observed at low redshifts, all the results are in relatively good agreement with current observational data.